\documentclass[letterpaper]{article} 
\usepackage{aaai24}  
\usepackage{times}  
\usepackage{helvet}  
\usepackage{courier}  
\usepackage[hyphens]{url}  
\usepackage{graphicx} 
\usepackage{natbib}  
\usepackage{caption} 
\usepackage{algorithm}
\usepackage{algorithmic}

\usepackage{subcaption}

\usepackage{newfloat}
\usepackage{listings}
\DeclareCaptionStyle{ruled}{labelfont=normalfont,labelsep=colon,strut=off} 
\floatstyle{ruled}
\newfloat{listing}{tb}{lst}{}
\floatname{listing}{Listing}
\title{Mechanic Maker: Accessible Game Development Via Symbolic Learning Program Synthesis}
\author{
    Megan Sumner\equalcontrib,
    Vardan Saini\equalcontrib,
    Matthew Guzdial\equalcontrib
}
\affiliations{
    University of Alberta, Alberta Machine Intelligence Institute\\

    Edmonton, Canada\\
    msumner@ualberta.ca, vardan1@ualberta.ca, guzdial@ualberta.ca
}

\begin{document}

\maketitle
\begin{abstract}
Game development is a highly technical practice that traditionally requires programming skills. This serves as a barrier to entry for would-be developers or those hoping to use games as part of their creative expression. While there have been prior game development tools focused on accessibility, they generally still require programming, or have major limitations in terms of the kinds of games they can make. In this paper we introduce Mechanic Maker, a tool for creating a wide-range of game mechanics without programming. It instead relies on a backend symbolic learning system to synthesize game mechanics from examples. We conducted a user study to evaluate the benefits of the tool for participants with a variety of programming and game development experience. Our results demonstrated that participants' ability to use the tool was unrelated to programming ability. We conclude that tools like ours could help democratize game development, making the practice accessible regardless of programming skills. 
\end{abstract}

\section{Introduction}
Developing a video game requires significant time and skill, with the complexity of games increasing over time \cite{VideoGameLength}. This limits who can engage with games as an artistic medium, as a tool for procedural rhetoric \cite{treanor2013account}, or as an education aide \cite{EducationalGames}. If the time and technical barrier of entry into game development could be reduced, game development would be more approachable, allowing access to game creation for those who have been historically excluded. However, this requires more accessible game development tools particularly focused on reducing the requirements of programming skills. 
Specifically, we anticipate that a tool for creating game mechanics, as a crucial component to video games, would be a useful first step towards this broader goal. 

To provide clarity of our meaning when we discuss a game mechanic (often shortened to mechanic), we use the definition from Zubek \cite{ElementsOfGameDesign}. Zubek defines a game mechanic as ``the basic activities of a game and the rules that govern them''. When we discuss rules, we refer to rules as they are represented by code in a game. Rules are implemented in a specific programming language and are executable by a game engine. A game engine is a software environment that can be used to develop a video game. When we discuss a game engine throughout this paper we are narrowing the focus to a group of rules provided from the backend that execute to create a playable game mechanic. The game engine provides usable rules to execute the game mechanic the user is creating. 

There are various tools that can help to create mechanics without the typical programming requirement, but they have limitations. Tools for game mechanic creation without typical programming take one of two forms: (i) visual programming, or (ii) pre-authored components with user-defined parameters. Visual programming allows users to create mechanics through visual elements instead of writing code in text. Visual programming tools include Scratch \cite{ScratchProgramming} and Unreal Engine Blueprints \cite{UnrealEngine}. While helpful, visual programming languages do not fully remove the technical barrier as they still require programming knowledge, just with a different interface. Pre-authored components have proven popular in commercial game development tools like Kodu Game Lab \cite{KoduGameLab} and Dreams \cite{Dreams}. Recombining pre-authored components to produce new mechanics removes the need for programming, but pre-authored components are limiting in the types of mechanics that the user can create.

In this paper, we present Mechanic Maker, a game mechanic creation tool which requires no programming knowledge. Instead, it works through a backend symbolic Machine Learning (ML) approach that creates code (i.e. program synthesis), based on a user demonstrating mechanics they want to exist. In this way, a user can define a variety of mechanics. For example, a user could create a character moving through keyboard input, an object spawning when another object gets to a certain position, a character jumping when space bar on the keyboard is pressed, and so on. We do not have pre-authored rule templates or otherwise limit users, and our backend symbolic learning approach to program synthesis (SLPS) can learn any deterministic sequence of rules from Markovian states \cite{guzdial2017game}. In effect, this example-based mechanic development paradigm allows users to describe the end goal they want for their mechanics, with Mechanic Maker doing the ``programming'' for them. 

To evaluate Mechanic Maker, we conducted a human subject study with users with a range of prior programming experience. Our results suggest that users find Mechanic Maker valuable for creating game mechanics, and that it is equally useful for programmers and non-programmers. This demonstrates the potential for expanding this methodology through intelligent tools to democratize game creation. Our contributions include Mechanic Maker itself, the underlying SLPS approach extended from \cite{guzdial2017game}, and the results of our human subject study.

\section{Related Work}

Mechanic Maker builds on a few different areas of games research including co-creative tools, autonomous rule generation and program synthesis.

\subsection{Co-Creative Tools}
Human-computer co-creativity involves a human and computer working together in the creative process \cite{Davis_2021}. Specifically related to game development there are many co-creative tools for developing levels for games \cite{DungeonGenerator,LodeEncoder}. Most approaches to co-creativity use search-based Procedural Content Generation (PCG) or another non-learning PCG approach \cite{DesignDrivenCoCreation}. Previous works have investigated machine learning approaches with explainable AI \cite{ExplainableAI}. However, such approaches tend to still assume familiarity with machine learning, which we avoid in an effort to reduce technical barriers. Machado et al. \cite{machado2019evaluation} have a co-creative tool for game creation that uses an AI-driven game development assistant to suggest existing mechanics from other games based on what the user has developed so far. This tool does not learn and adjust based on input and uses existing rules instead of creating new ones like Mechanic Maker.

There are two prior examples of co-creative tools that learn and adjust their actions based on user feedback \cite{Guzdial_2019,halina2022threshold}. Guzdial et al. \cite{Guzdial_2019} developed a tool called Morai Maker which takes turns with users to create levels and attempts to learn the style of the human user. Halina and Guzdial \cite{halina2022threshold} developed a rhythm game generator called KiaiTime that similarly attempts to learn the design style of a human user. Our proposed tool, Mechanic Maker, does not involve a turn-taking interaction like these two tools, instead continually updating learned rules based on user demonstrations. Mechanic Maker also focuses on game rules instead of game levels.

Kruse et al. \cite{gamedesign} presented a human-in-the-loop game design study to determine the use of PCG tools for professional game designers. Their results showed that there is still work to be done to get game developers interested in AI assisted tools. This is due to the technical difficulty in using existing PCG tools in games and the unreliable results that the tools can provide. We aim for our tool to address the problem of technical difficulty as it removes the need for coding or parameter tuning.

Guzdial et al. \cite{guzdial2017game} employed their Engine Learning algorithm to learn the rules from three existing games from gameplay video and then attempted to combine rules from these existing games to generate new rules. While we based our SLPS approach on their Engine Learning algorithm, which we discuss further below, we extend this algorithm in order to make it appropriate for real time interaction. 

\subsection{Autonomous Rule Generation}

Autonomous generation stands at the other side of the spectrum from co-creative approaches, in which a system generates content without any human interaction outside of starting the process \cite{deterding2017mixed}. The majority of existing rule generation work has relied upon autonomous generation rather than anything with a human in the loop. We can split autonomous rule generation work into two groups, (i) search-based PCG and (ii) world models. 

In search-based PCG, a generator searches over a space of possible rules to produce ones that are good according to a human-authored evaluation function.
Search-based rule generation dates back to the early 1990s \cite{pell1992metagame}. The most common approach requires authoring possible rules effects, which can then be selected based on a search-based optimization \cite{togelius2008experiment,sorochan2022generating}. Cook et al. and the recent follow up by Gonzalez et al. \cite{cook2013mechanic, MechanicMiner} employed code reflection to identify possible public variables which could then appear in different rule effects. Their work, similarly to our own, did not rely on pre-authored rule effects, though we instead learn them from user examples.

With world models, a machine learning model is fed with examples of rules from a real game, and then attempts to approximate them either explicitly or implicitly \cite{WorldModels, GameGAN, genie}. They train on a massive amount of data to approximate a particular game environment. Unlike our approach, world models do not generally learn explicit code, instead relying on fuzzy neural predictive models.
Guzdial and Riedl \cite{guzdial2021conceptual} represents the only example, to the best of our knowledge, of combining world models and autonomous rule generation.

\subsection{Program Synthesis}
Program synthesis represents the task of automatically generating programs to accomplish some task \cite{ProgramSynthesisSurvey}. It is not typically applied to game development, though some prior work exists within the domain of games \cite{kreminski2021opportunities}.
Similar to Mechanic Maker, Medeiros et al. \cite{ProgrammaticStrategies} synthesize programs that represent strategies based on player behaviors. But they do not apply program synthesis to generate game content.

Yang et al. \cite{RLSynthesis} apply program synthesis for approximating unseen parts of partially observed environments. While Mittelmann et al. \cite{AutomatedMechanisms} design game theoretic environments based on optimal strategies using program synthesis. Both of these prior approaches apply program synthesis to produce dynamic information about game-like environments. In comparison, our program synthesis approach relies on human input and feedback to design playable video games. 

Simplified game engines can be used to reduce the technical barrier of game development \cite{SGE}. The Gemini game generator \cite{Gemini2018} creates games using a predefined set of rules. Gemini synthesizes these rules based on a provided meaning by a user in a domain-specific language (DSL). While this tool doesn't require directly programming a game, it does require specialized technical knowledge in terms of authoring intended meanings in its DSL.

\section{Mechanic Maker}

Mechanic Maker is a game development tool initially developed by Saini and Guzdial \cite{Saini_Guzdial_2020}. The Mechanic Maker tool is split into two different sections:

\begin{enumerate}
    \item The Symbolic Learning Program Synthesis (SLPS) backend. This backend interfaces with the game mechanic editor to learn from the user inputs.
    \item The game mechanic editor frontend. This frontend allows users to create game mechanics and receive predictions from the SLPS backend.
\end{enumerate}

\subsection{Symbolic Learning Program Synthesis}

The SLPS backend extends the Engine Learning algorithm by Guzdial et al. \cite{guzdial2017game}. We made the following changes to the Engine Learning algorithm from Guzdial et al. First, the original algorithm assumes an unchanging sequence of game frames from a gameplay video. We instead changed the algorithm to run iteratively every time there was a new frame, initializing with the last learned engine instead of an empty engine. Second, we altered the types of Facts, breaking their ``Spatial'' Fact type into a PositionX and PositionY Fact type, and removing the CameraX Fact type. We did this to learn more nuanced rules for the former and as there is no camera movement for the latter. An example of the format of one of the learned rules is shown in the appendix. 
Third, we also set the Engine Learning loop to a maximum number of ten iterations based on initial testing, rather than requiring it to run until it has a perfect prediction. We needed this as the Engine Learning algorithm assumes no hidden information or randomness, and it could fail if the user introduced either. In such cases, we return the closest learned Engine after termination. With these changes, our SLPS backend can take in new frame content and output an engine as a sequence of rules at runtime that works with the frontend game editor to represent a game's mechanics. 

%

We show our SLPS algorithm in Algorithm \ref{alg:algorithm}. The input is a sequence of frames. Each frame is defined as the set of inputs the user defined and a set of objects, which are themselves defined by their image, x and y coordinates, and velocity in the x and y direction. SLPS translates each frame into relevant fact types: animation facts (that track image usage), velocity x facts, velocity y facts, position x facts, position y facts, variable facts (tracking input information, each fact tracking one button being pressed or not), relationship x facts (tracking x-dimension relationships between objects), relationship y facts, and empty facts (which are necessary to handle appearing and disappearing rules). 

For each frame in this fact representation, SLPS attempts to predict the next frame with a given current engine (defined as a sequence of learned rules). This is executed in Algorithm \ref{alg:algorithm} inside the \emph{Distance()} function. Initially this engine is empty. We populate the engine with rules using a search-based process, shown in Algorithm \ref{alg:algorithm} as \emph{EngineSearch()}. This function optimizes the engine by adding, modifying, and removing rules.

Adding a rule is a search operator that takes a pair of unmatched facts between the frames and creates a rule. This rule is created by setting all the facts of the current frame as the conditions that must be true for the rule to fire. It also adds the unmatched facts in the form of the pre-effect fact and post-effect fact. Modifying rules is a search operator that simplifies an existing rule by taking the union of the facts in a rule’s condition and in the current frame as the modified rule's conditions. This allows for rules to generalize over time. Removing rules is a search operator that deletes rules. In this way, the engine is optimized by using the Hellman’s metric to minimize the number of differences between the frames in terms of facts. 

If we update the engine being used, shown in Algorithm \ref{alg:algorithm} as \emph{UpdatedSuccessfully()}, we restart the predict process from the first frame with the new engine now being used. This is done until a prediction within the threshold is reached, shown in Algorithm \ref{alg:algorithm} as the \emph{Predict()} function, or we have searched through 10 neighbouring engines, at which point we return the engine that minimized the Hellman’s metric.

\begin{algorithm}[tbh]
\caption{SLPS Engine Learning Algorithm}\label{alg:algorithm}
\begin{algorithmic}
\STATE input: A sequence of continuous and valid frames of size $f$ and threshold $\theta$
\STATE output: $Engine$ $engine;$
\WHILE{$True$ (Until runtime stopped)} 
    \STATE $e \gets new Engine();$
    \STATE $cF \gets frames[0]$
    \STATE $MaxIterations \gets 10$
    \WHILE{$i \gets 1$ to $len(frames)$}
        \STATE \emph{Attempt to learn an engine within the given MaxIterations}
        \WHILE{$iterations \leq MaxIterations$}
            \STATE \emph{Check if this engine predicts within the threshold.}
            \STATE $frameDist \gets Distance( e, cF, i+1)$;
            \IF{$frameDist \leq \theta$}
                \STATE $cF \gets Predict(e, cF, i + 1)$
                \STATE $break;$
            \ENDIF
            \STATE \emph{Update engine and start parse over;}
            \STATE $e \gets EngineSearch(e,cF, i+1)$
            \IF{UpdatedSuccessfully(e, cf, i+1)}
                \STATE \emph{Reset frames and start again}
                \STATE $i \gets 0;$
                \STATE $cF \gets frames[0];$
                \STATE $iterations \gets 0;$
                \STATE $break;$
            \ENDIF
            \STATE $iterations \gets iterations+1;$
        \ENDWHILE
        \STATE $i \gets i+1;$
    \ENDWHILE
\ENDWHILE
\end{algorithmic}
\end{algorithm}

\subsection{Game Mechanic Editor} 
\begin{figure}[tbh]
\centering
\begin{subfigure}[b]{0.35\textwidth}
    \includegraphics[width=\textwidth]{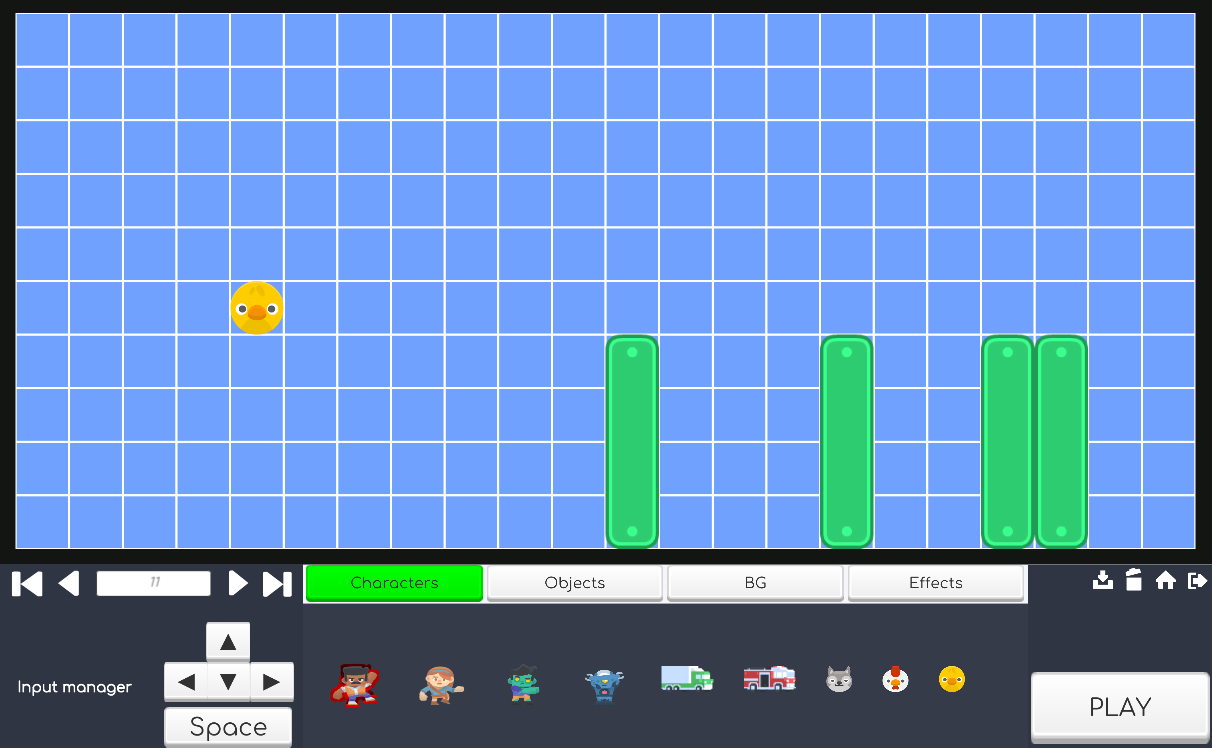}
    \caption{Frame Editor for frame 0}
\end{subfigure}
\begin{subfigure}[b]{0.35\textwidth}
    \includegraphics[width=\textwidth]{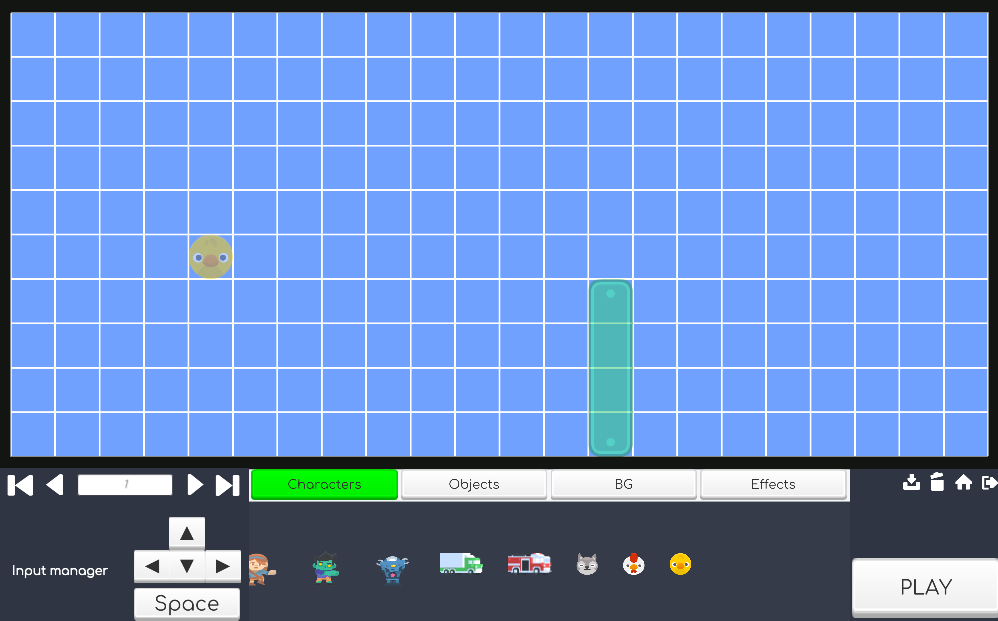}
    \caption{SLPS Prediction}
\end{subfigure}
\begin{subfigure}[b]{0.35\textwidth}
    \includegraphics[width=\textwidth]{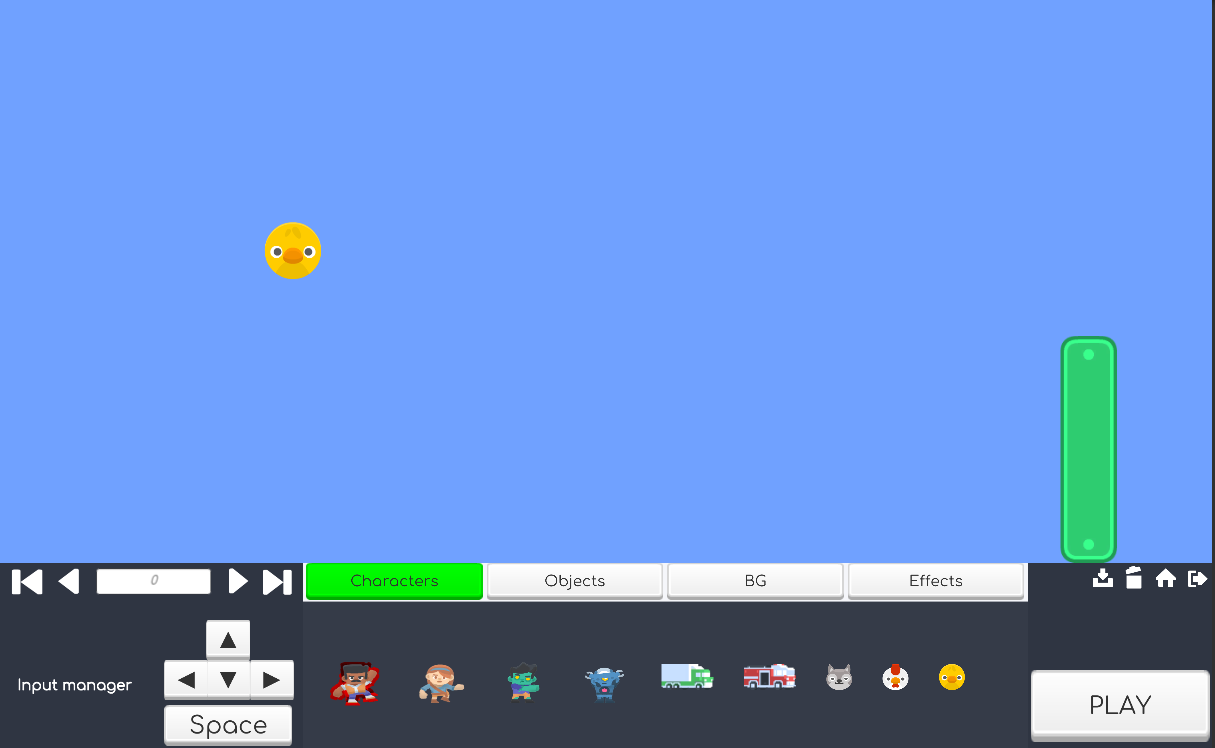}
    \caption{Play Mode}
\end{subfigure}
\caption{The Mechanic Maker editor. (a) The user defines frames of their game by placing objects on a grid, (b) the SLPS backend attempts to learn the underlying rules suggested by the changes across the frames, and (c) the user can test the learned rules in real-time via the Play Mode.}
\label{fig:UserInterface}
\end{figure}

There are three main components to our game mechanic editor frontend:

\begin{enumerate}
  \item The frame editor shown in Figure \ref{fig:UserInterface}.a.
  \item Predictions from the SLPS backend, with an example shown in Figure \ref{fig:UserInterface}.b.
  \item The Play Mode shown in Figure \ref{fig:UserInterface}.c.
\end{enumerate}

The frame editor in Figure \ref{fig:UserInterface}.a is the main point of interaction in our tool. We present the controls of our game mechanic editor at the bottom of each sub-figure in Figure \ref{fig:UserInterface}. Here the user can specify the frame to modify and what inputs the player gives to demonstrate the desired mechanics. The user specifies the intended behaviours for their game mechanics by demonstrating them across these frames. For example, if a user wants to have an object moving to the right, they would add it to the grid for frame 0 then place the same object one position to the right in the next frame. 
The SLPS backend, discussed above, will then learn that the object should move to the right when the game is played. The frames allow the user to specify the desired effects of the game mechanics and represents the training data for the SLPS backend to learn what mechanics the player wants in the game.

As demonstrated in Figure \ref{fig:UserInterface}.b, when the user moves to a new frame they see a semi-transparent ``ghosting'' of the SLPS backend's current predictions based on its current learned mechanics. At first, when nothing has been learned, this will simply predict the prior frame (assuming nothing changes). However, as mechanics are learned, the tool will reflect them in its prediction. The user can choose to accept the prediction or ignore it, and edit the frame as they see fit.

At any time, the player can test the current learned rules in real-time by pressing the play button in Figure \ref{fig:UserInterface}.c. This allows them to verify that the game is working as intended, as shown in Figure \ref{fig:UserInterface}.c. If not, they can keep iterating and adding more frames to correct the SLPS backend's learned mechanics.

\section{Human Subject Study}

In this section we cover the setup of our human subject study. We ran our human subject study to evaluate our Mechanic Maker tool and investigate the hypotheses listed below. We obtained ethics approval via the University of Alberta Research Ethics Board (REB), Pro00102469.

\subsection{Hypotheses}

We focused on testing three hypotheses in our user study. The first hypothesis is that programming experience is not required to use our tool effectively. This hypothesis would help us determine if our tool is capable of lowering the technical barrier for creating game mechanics, which is the primary focus of this research.
We next wanted to test that Mechanic Maker has value as a game development tool. Testing the value of the tool for game development is more difficult to measure objectively due to a lack of existing measures for game development quality. Thus we split this into two different hypotheses related to game development value and measure them separately. 
The second hypothesis is the \emph{usability} of this tool. We define usability in this work as the extent to which the tool was able to help each participant create the game mechanics they desired.
The third hypothesis is \emph{participant enjoyment} when using the tool. We want to know if the tool was enjoyable for the participants to use, which will help us determine the overall user experience for creating game mechanics.
Support for these hypotheses would demonstrate the potential for achieving our goal of democratizing game development by continuing to build similar tools or iterating upon Mechanic Maker. 

\subsection{Procedure}
Our study consisted of an hour long process broken into four parts. Each part had a time limit. If the user did not complete a part within the time limit, a facilitator asked them to move on to the next part.
We did this for consistency, to respect participants' time, and as we found longer periods with the tool did not improve outcomes. 

The first part of the study was a step-by-step tutorial to provide users with an understanding of Mechanic Maker. It involved recreating the game mechanics for a game called Sokoban where the objective is for the player to move a crate onto a goal to beat the level. Our tutorial was a simplified version of this that introduced the participant to player movement and pushing the crate to the right. Participants were given a reference video of one of the study personnel walking through how to create the Sokoban mechanics with our tool, which we iterated on based on a pilot study of four individuals. We allocated 25 minutes to this portion of the study as the participants were still learning how the tool worked and how to interact with the UI.

The second part of the study involved asking the participants to replicate a simplified version of the game mechanics from Flappy Bird. Flappy Bird consists of pipes moving to the left with the player controlling a bird, attempting to dodge the pipes for as long as possible. We again gave participants access to a reference video. However, in this case, we only gave them video of the final game mechanics running, not of the steps needed to recreate them with Mechanic Maker. The video showed a simplified version of Flappy Bird involving only one pipe that would go to the end of the screen and then teleport to the other side to keep moving to the left, and a bird that would continually fall, and jump upwards when the space bar was pressed. A screenshot of the Flappy Bird example is shown in Figure \ref{fig:UserInterface}.c. There were no Game Over elements or scoring involved in this process. The participants had 15 minutes for this part of the study.
If participants could successfully recreate the simplified Flappy Bird regardless of programming ability that would support our first hypothesis that Mechanic Maker did not rely on programming ability. It would also support our second hypothesis around the usability of the tool.

The third part of the study gave participants the opportunity to create game mechanics of their choice. They were given 15 minutes to come up with game mechanics and implement them in Mechanic Maker. They were asked to take what they had learned from the previous exercises to ``create a simple game''.
If participants were able to successfully create a wide variety of games, that would support our second hypothesis of the value of Mechanic Maker for game development in terms of usability. Further, if they enjoyed this process, this would provide support for the third hypothesis of participant enjoyment.

The final part of our study was a survey that took 5 minutes to complete. We cover the survey in the next subsection. 
We included the survey in order to collect self-reported and demographic information related to our hypotheses.

\subsection{Survey Design}

The survey was composed of 19 Likert scale questions, three short answer questions, and ten demographic questions. While we only gave participants five minutes for the survey, we found this to be ample time during our pilot studies.
The first section of the survey used a four point Likert scale with 1 being not at all true, 2 being not true, 3 being true and 4 being very true. The reason for this was to attempt to minimize the neutrality bias. We adapted the first nineteen questions from the Intrinsic Motivation Inventory (IMI) \cite{IntrinsicMotivationInventory} due to the lack of a validated survey for co-creative tools. These questions were centered around our second and third hypotheses, asking users about their usability and enjoyment of the tool in each part of the study. We acknowledge that there are other surveys more related to usability and that IMI is not used as a complete measure, but adapting our questions from IMI was sufficient for our initial Mechanic Maker study. 

The second section included three short answer questions for feedback around what users would change and overall thoughts around the tool.
The final section asked demographic questions, including the users programming experience and game development experience, shown in Table \ref{table:Participants}, which we used to evaluate our first hypothesis. We include all original survey questions in the appendix.

\section{Results}
In our human subject study we collected two types of information. First, from the survey we collected self-reported quantitative, qualitative, and demographic information. Second, we logged all major Mechanic Maker events, all final games, and all of the frames produced by users. In the below subsections we report our results related to our three hypotheses.

\subsection{Participants}
We had 16 participants with varying programming and game development backgrounds take part in our study, as shown in Table \ref{table:Participants}. We had 8 participants between the ages of 18-25, 7 between the ages of 25-35 and 1 between the ages of 35-45. We had 11 male, 3 female and 2 non-binary or other participants. This is not an equitable distribution, but shows similar gender bias to what is seen in the games industry and the tech industry broadly \cite{statistaGameDev}. 

\begin{table}[tbh]
\centering
\begin{tabular}{c l l } 
 \hline
 Participant ID & Programming & Game Development\\ [0.5ex] 
 \hline
 1 & Limited & Limited\\ 
 2 & Expert & Moderate\\
 3 & Limited & Limited\\
 4 & Expert & Expert\\
 5 & Expert & Moderate\\
 6 & Expert & Limited\\
 7 & Moderate & Moderate\\
 8 & None & None\\
 9 & Moderate & Limited\\
 10 & Expert & Moderate\\
 11 & Expert & Moderate\\
 12 & Expert & Moderate\\
 13 & Limited & None\\
 14 & Limited & Limited\\
 15 & None & None\\
 16 & Limited & Limited\\
 \hline
\end{tabular}
\caption{Experience of participants in programming and game development.}
\label{table:Participants}
\end{table}

\subsection{Frame Error}

To measure the success of users at replicating the reference games, we introduce a metric called \emph{Frame Error}. As a reference point, we drew on the frames that were used to create the example games in the tutorial videos for both Sokoban and Flappy Bird. These examples frames were authored by study personnel. For each participant, we measured how well the learned mechanics from that participant could predict each example frame given the prior example frame. We defined \emph{Frame Error} using the Hellman's Metric to determine how close a predicted frame was to the true next frame.
This metric is useful for determining success at replicating the reference games as it gives us an error equivalent to the dissimilarity of the participant's learned mechanics from the intended mechanics. 

\subsubsection{Baseline}
For our baseline, we used the previous example frame with no changes as the prediction, measuring the difference with Hellman's metric as above. This is equivalent to the prediction of an empty engine, and was found to significantly outperform specialized frame prediction models with low training data in prior work \cite{guzdial2017game}. 

\subsubsection{Frame Error Results}
\begin{figure}[tbh]
    \centering
    \begin{subfigure}{0.4\textwidth}
        \includegraphics[width=\textwidth]{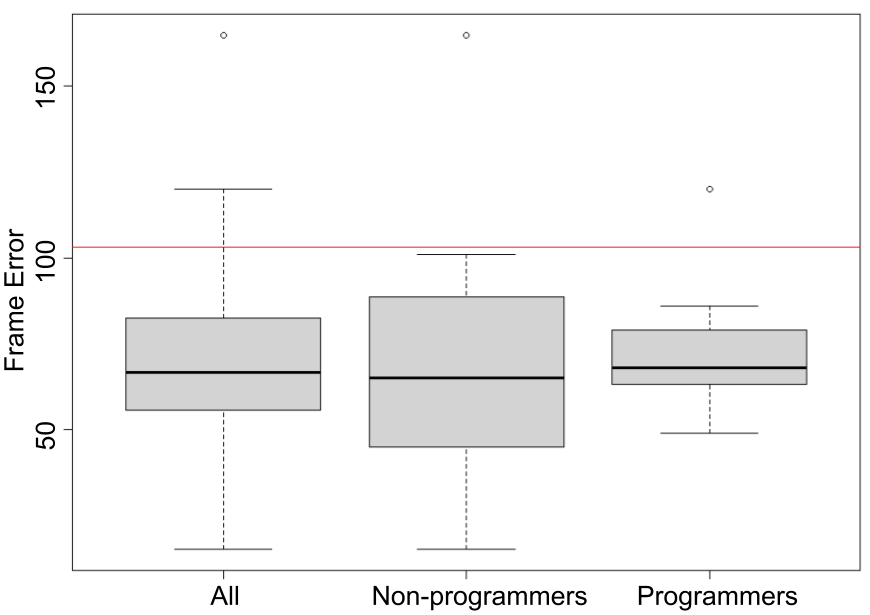}
        \caption{Sokoban Activity}
    \end{subfigure}
    \begin{subfigure}{0.4\textwidth}
        \includegraphics[width=\textwidth]{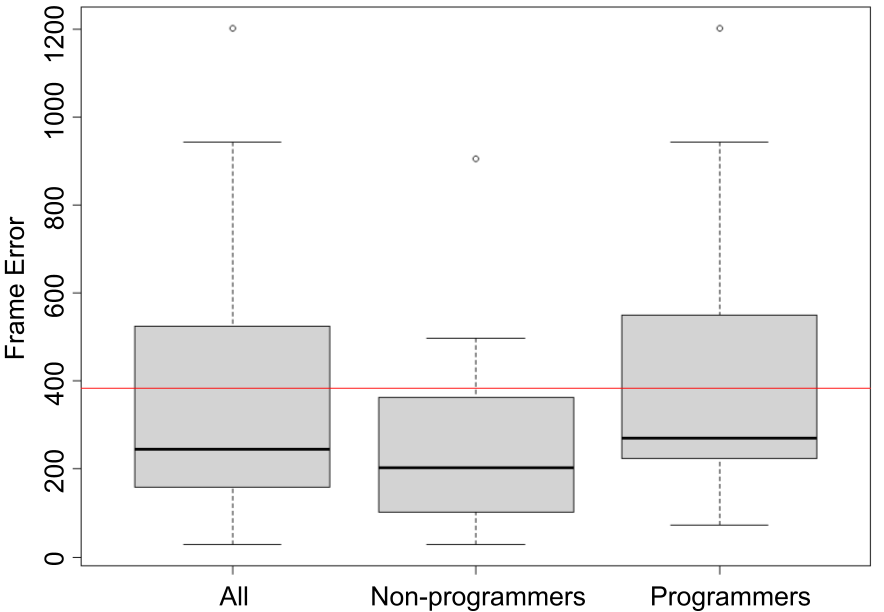}
        \caption{Flappy Bird Activity}
    \end{subfigure}
    \caption{Frame error grouped by programming experience for the (a) Sokoban and (b) Flappy Bird activities. From the survey results None and Limited programming were merged into the non-programmer category and Moderate and Expert programming experience were merged into programmer. The All box plot shows all programming experience combined. The red line marks the performance of our baseline.}
    \label{fig:FrameError}
\end{figure}

The results are shown in Figure \ref{fig:FrameError}. For both of these plots, we reduced the programming experience groups from our survey participants shown in Table \ref{table:Participants} from four groups (None, Limited, Moderate, Expert) to two (Non-programmers and Programmers). This was done to simplify the box plots as we are only looking at whether lower programming experience led to any difference in the results compared to higher levels of programming experience. 

As shown in Figure \ref{fig:FrameError}.a, for the Sokoban part, programming experience did not have an impact on the frame error at all, and both values were below the baseline value (the red line). In the Flappy Bird example in Figure \ref{fig:FrameError}.b, the same is true. The median values of frame errors are lower than the baseline, indicating that the majority of participants led to learning useful rules. The non-programmer distribution is actually moderately lower than the programmer distribution. These results support our first hypothesis.

We also ran a Pearson correlation test for the frame error for both the Sokoban and Flappy Bird parts using the four choice options for programming experience. We found that in both parts there was no correlation between frame error and programming experience. Sokoban had a correlation value of $0.11$ and Flappy Bird a correlation value of $0.21$, with neither being significant.


\subsection{Free Play Analysis}

\begin{figure}[tbh]
    \includegraphics[width=.235\textwidth]{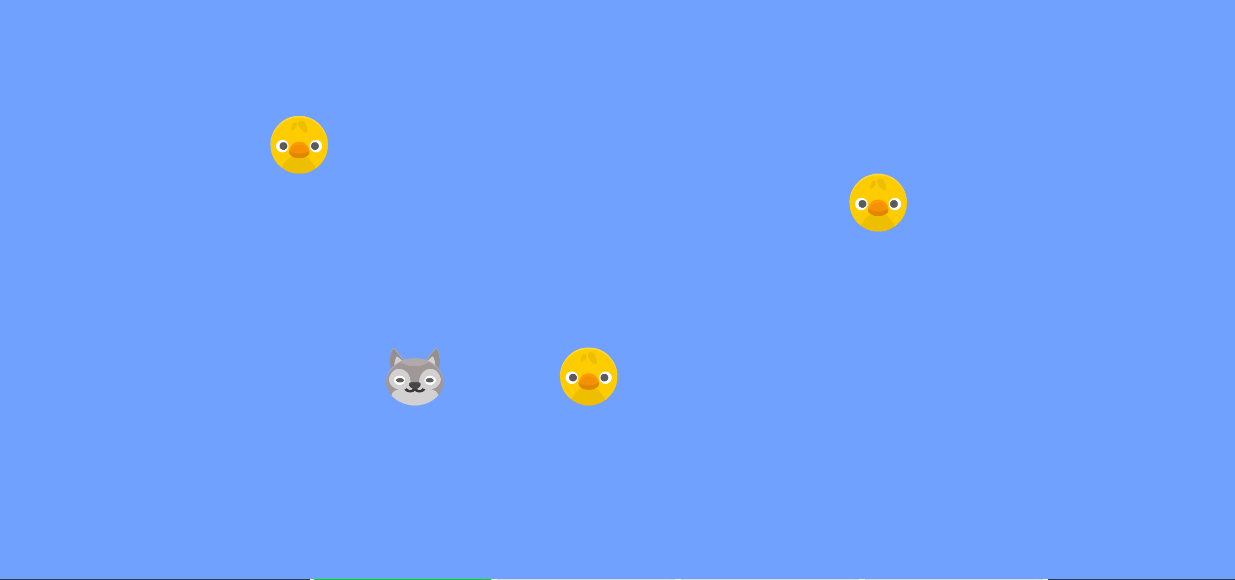}\hfill
    \includegraphics[width=.235\textwidth]{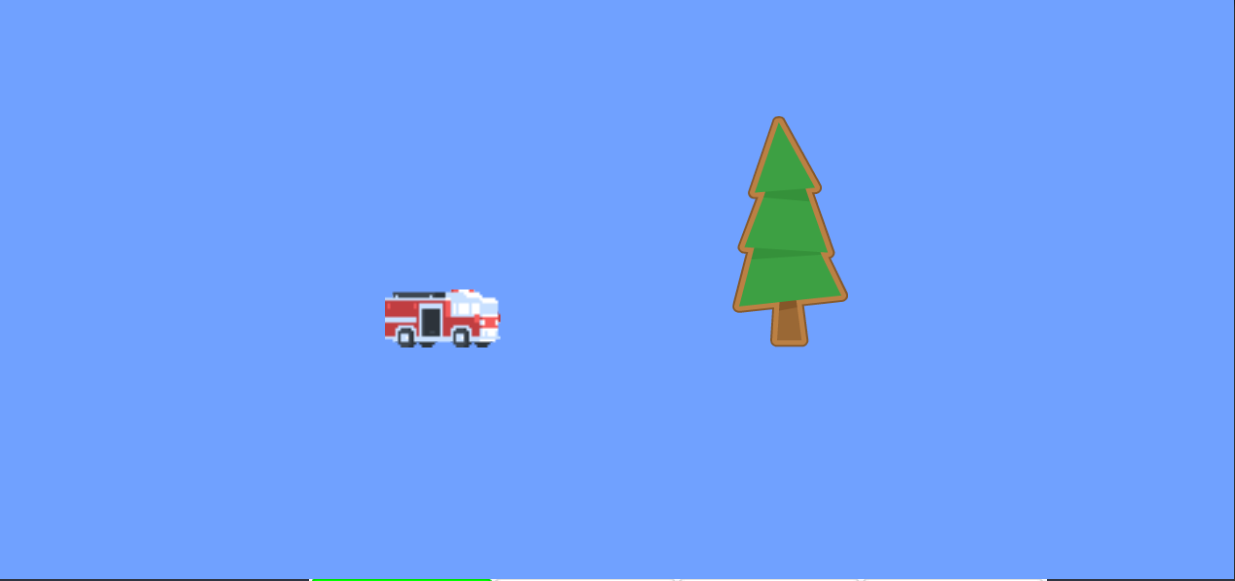}\hfill
    \includegraphics[width=.235\textwidth]{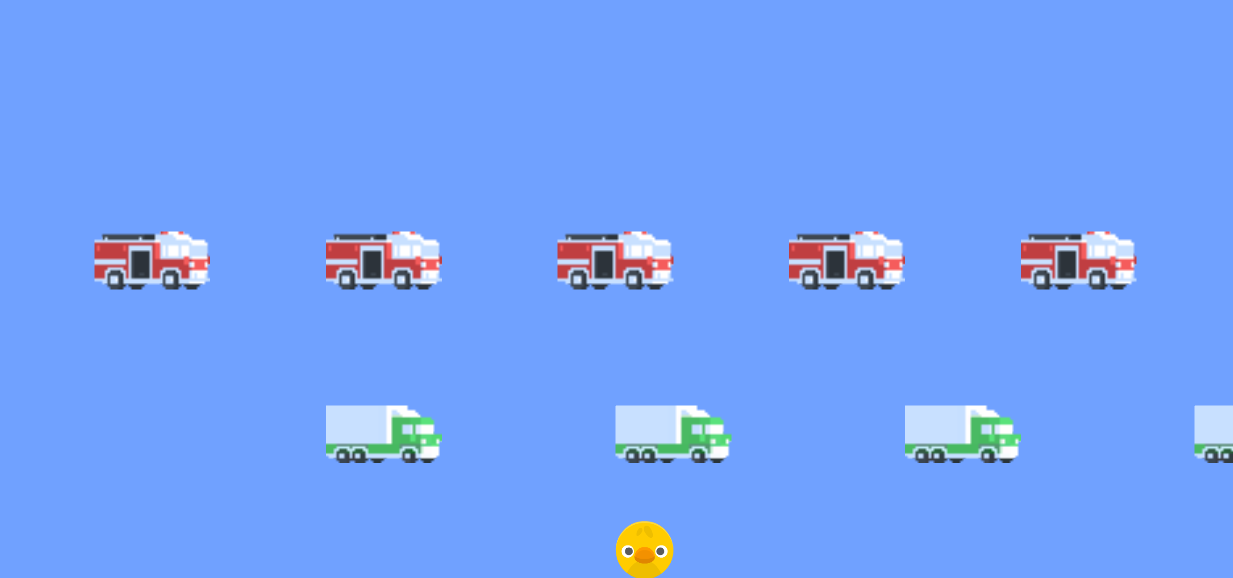}\hfill
    \includegraphics[width=.235\textwidth]{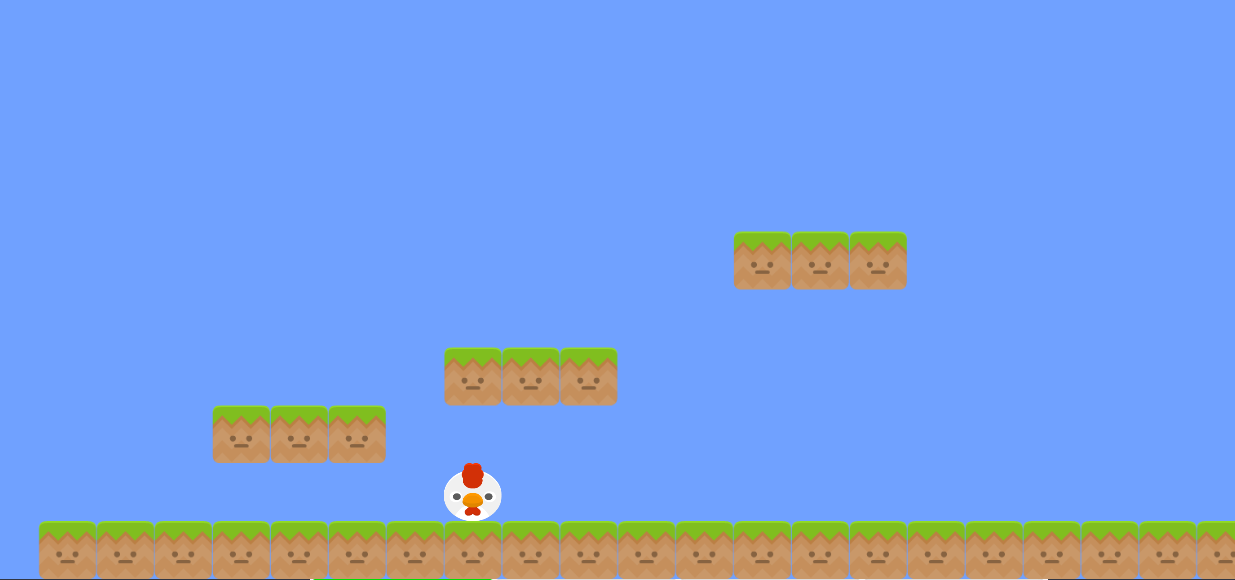}\hfill
    \includegraphics[width=.235\textwidth]{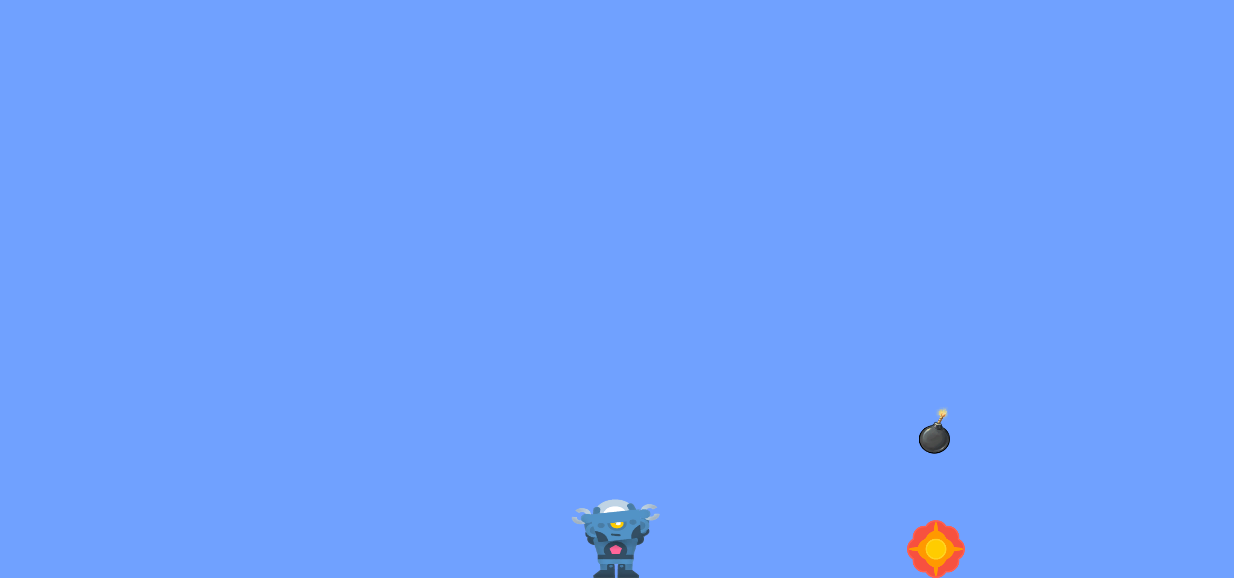}\hfill
    \includegraphics[width=.235\textwidth]{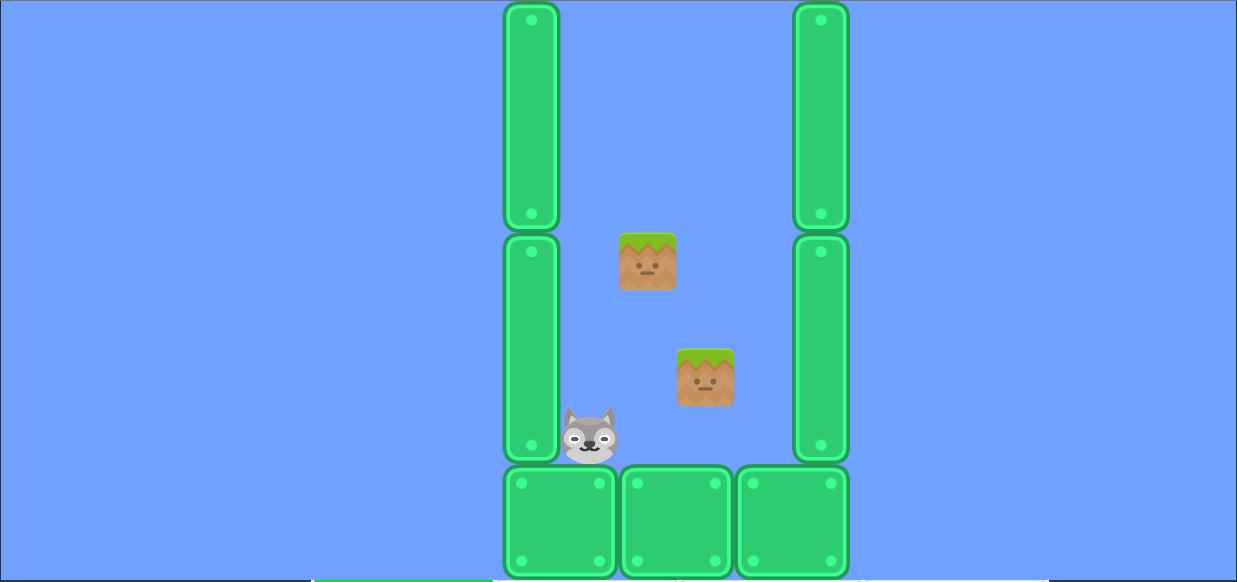}
    \caption{Example game outputs from the Free Play portion of the user study.}
    \label{fig:FreeplayExamples}
\end{figure}

Given that there was no ground truth for the Free Play part of the study, we cannot calculate frame error. Instead, we determine whether participants were able to make a variety of game mechanics, or if participants were limited in terms of making game mechanics like those in Sokoban and Flappy Bird. Ideally we would have some measure of how different the games are from one another to demonstrate that Mechanic Maker can create a variety of games. Since this is difficult to quantify, we use standard deviation metrics as a proxy for the variation. Using this metric we found the standard deviation to be $ 67.06 \pm 48.28 $ and $ 10.13 \pm 5.74 $ for the number of frames and number of mechanics created, respectively. Given the large standard deviation values, in the number of frames in particular, this provides some support for there being a large variety of games, which can also be seen in the six selected games shown in Figure \ref{fig:FreeplayExamples}. This provides some evidence to support our second hypothesis, that Mechanic Maker can create a variety of different game mechanics.


\subsection{Mixture Model}
\begin{figure}[tbh]
\centering
\includegraphics[scale=0.3]{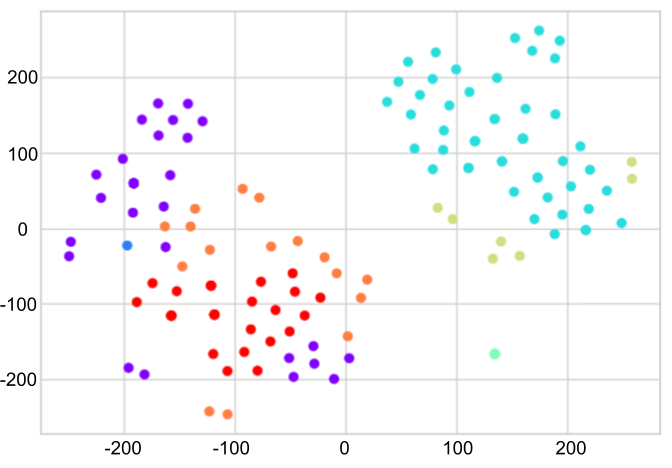}
\caption{T-distributed Stochastic Neighbor Embedding (TSNE) to visualize the twenty dimensions of the GMM as two dimensions. This Figure displays a visualization of our GMM as seven clusters.}
\label{fig:TSNE}
\end{figure}

In an attempt to better understand the variability of the free play results we created a Gaussian Mixture Model (GMM) from the user study data. This was inspired by previous attempts to better understand design tasks through clustering \cite{alvarez2022toward}. We formatted each rule using a one-hot encoding of the fact type (velocity, position, animation) and the relevant value information of the fact for both the pre-effect and post-effect. We also included a count of how many of each type of condition (by fact type) was used in each rule. From there, we used a GMM to group the different rule representations together into clusters. We used the elbow method to determine the number of optimal Gaussians for our data and found that 7 clusters created a good representation. A visualization of the distribution of Gaussians is shown in Figure \ref{fig:TSNE}. Each point represents a learned rule, and each colour represents the Gaussian that each rule largely falls within. With this information we can describe the means of each cluster to get a better understanding of the groupings. In the top right of Figure \ref{fig:TSNE} there are two clusters — clusters 2 and 4. These two clusters are associated with pre-effects and post-effects with velocities in the y direction, while all the other clusters are related to velocities moving in the x direction. The isolated cluster 3 is related to no movement in either the velocity x or y, encompassing rules that include collision information. This demonstrates a wide variety of rules learned during the free play portion, with a bias towards velocity rules. 

\subsection{Survey Results}

\begin{figure}[tbh]
\centering
\includegraphics[scale=0.32]{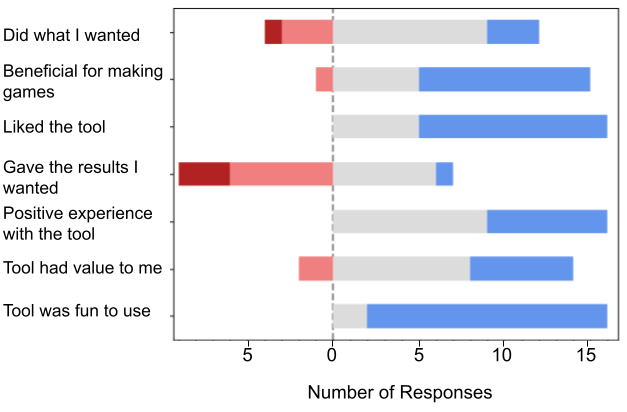}
\caption{Survey results for the tool in general.}
\label{fig:SurveyGeneral}
\end{figure}

\begin{figure}[tbh]
\centering
\includegraphics[scale=0.32]{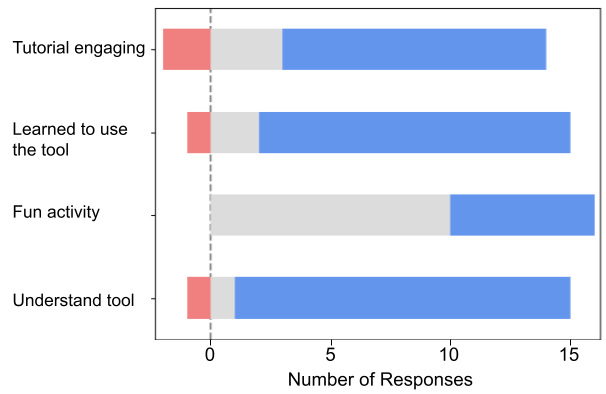}
\caption{Survey results for the tool when doing the Sokoban activity.}
\label{fig:SurveySokoban}
\end{figure}

\begin{figure}[tbh]
\centering
\includegraphics[scale=0.32]{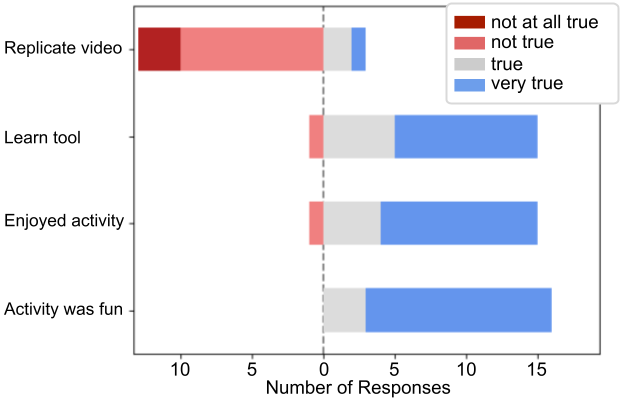}
\caption{Survey results for the tool when doing the Flappy Bird activity.}
\label{fig:SurveyBird}
\end{figure}

\begin{figure}[tbh]
\centering
\includegraphics[scale=0.3]{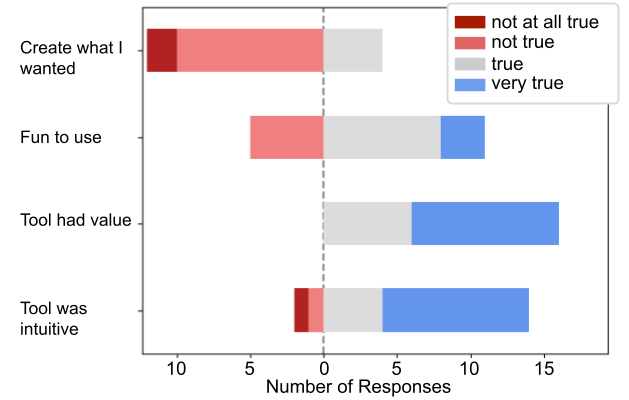}
\caption{Survey results for the tool when doing the Free Play activity.}
\label{fig:SurveyFreeplay}
\end{figure}

Figures \ref{fig:SurveyGeneral} to \ref{fig:SurveyFreeplay} show the relevant participant responses from the user study survey questions. We cover the survey results in three main categories.

\begin{enumerate}
  \item The enjoyment of using the tool — Did players enjoy using the tool?
  \item The usability of the tool — Were participants able to get what they wanted out of the tool?
  \item The value of the tool — Did the participants think this tool provided benefit to them?
\end{enumerate}

The results for enjoyment of the tool were largely positive. In Figure \ref{fig:SurveyGeneral} participants agreed that they liked the tool and that the tool was fun to use. In the individual portions of the study, the participants mostly found the tool fun to use.

For usability of the tool, the results in Figures \ref{fig:SurveyGeneral}, \ref{fig:SurveyBird} and \ref{fig:SurveyFreeplay} indicate that there is some issue for the participants achieving the results they wanted. This can be observed in the questions \emph{General - Gave the results I wanted}, \emph{Bird - Replicate video}, and \emph{Free Play - Create what I wanted}. As shown in our frame error analysis, participants were able to use Mechanic Maker effectively to create the appropriate rules when compared to the example games for Sokoban and Flappy Bird. This indicates they are generally achieving the desired effects, but that there is some friction in the interaction with the tool which we hope to address in future work.

The survey questions related to the value of the tool indicate an overall perception that Mechanic Maker in general is beneficial for making games and has value to the participants. It is promising to see that in Figure \ref{fig:SurveyFreeplay} participants felt that the tool had value. This supports our second hypothesis that Mechanic Maker provides benefit as a game development tool.

\section{Discussion}
In our survey we asked users for feedback on the tool to get a sense of what stood out. Most of the feedback was positive and indicated that Mechanic Maker had potential for game development. A couple handpicked quotes from the short-answer portion include ``I think this is a unique idea and approach to letting people make games. I feel like there is a great amount of potential for this tool, especially in teaching the basics of game design.'' and ``...It is also impressive in terms of being a `democratizing' tool i.e. allowing access to game-dev to anyone. I believe that tools like this are going to be a big part of the future of computing; good work!'' These quotes provide support for future iterations of the tool making game development more accessible.

\subsection{Ethical Statement}

Using AI tools to automate different areas of game development can cause concern in the games industry for a variety of different reasons. Generative AI tools including ChatGPT and Dall-E have recently sparked debate due to copyright issues. These tools are based around models that require massive amounts of training data, which can lead to datasets including works without a creator's consent. Our tool instead only uses data provided by the user. Another major area of ethical concern with AI is the loss of creative jobs including game art, voice overs, writing and other areas of a game \cite{Vimpari_2023}. Mechanic Maker does reduce the programming requirement for game development, which could lead to potential job losses in technical positions. However, this tool is focused on newcomers to game development. The emphasis is on reducing the technical barrier to allow more people to create games. Even though the focus is not on industrial applications, there is always a risk for new tools like this to lead to potential job losses. Regardless, since our tool is a collaborative effort between the AI and the user, there will always be the need to have people work with this tool co-creatively. 

\subsection{Future Work}
We identify that Mechanic Maker tended to perform best with fewer frames and objects. In future iterations, we hope to improve the performance of the tool.
Currently, we start every game creation session with a blank engine. However, we now have the results from the participants of this study, and so we hope to develop approaches to adapt knowledge from earlier learned engines to help better approximate a current game more quickly. Our current plan to improve the frame prediction is to use the GMM we've created to cluster mechanics. 
In particular, we anticipate a case-based reasoning or interpolation strategy to approximate new rules by querying the GMM may be effective at speeding up the rule learning process.


\section{Conclusion}
Game development is a complicated, technical field that traditionally requires significant programming skills. We propose Mechanic Maker, a new tool for game development that has the potential to eliminate, or at least mitigate, the necessity of programming to create game mechanics. Our user study shows that Mechanic Maker can be used to create intended games equally well between programmers and non-programmers. The study also provided an overall positive response to the tool. We believe Mechanic Maker has the potential to make game development more accessible and to democratize the field.

\section*{Acknowledgements}

This work was funded by the Canada CIFAR AI Chairs Program, Alberta Machine Intelligence Institute, and the Natural Sciences and Engineering Research Council of Canada (NSERC).

\bibliography{aaai24}
\newpage
\clearpage
\section{Appendix}
For more details on Mechanic Maker, we provide the full list of survey questions and an example rule output.
\subsection{Survey Questions from the User Study}
\begin{enumerate}
    \item I was able to use the tool to do what I wanted.
    \item I did not find the tool beneficial for making games.
    \item I liked the tool.
    \item I had a largely negative experience using the tool.
    \item The tool did not let me get the results I wanted.
    \item I believe that using this tool could be of some value for me. 
    \item This tool was fun to use.
\end{enumerate}

The final twelve Likert questions were broken into three groups of four questions, each group referencing one of the three game creation parts from the study:

\begin{enumerate}
    \item I found the Sokoban tutorial engaging.
    \item Following the tutorial helped me learn how to use the tool.
    \item I thought following the Sokoban tutorial was a very boring activity.
    \item I did not understand how the tool worked from the tutorial.
    \item I found it difficult to replicate the Flappy Bird example.
    \item I was able to learn to use the tool through this activity. 
    \item I enjoyed doing this activity very much.
    \item I did not have fun doing this activity.
    \item I was able to create what I wanted with the tool in Free Play
    \item Using the tool in Free Play was frustrating.
    \item I would be willing to do this activity again because it has some value for me.
    \item I found the tool too complicated to make what I wanted.
\end{enumerate}

Following these Likert questions we asked three short answer questions to identify additional qualitative information related to our hypotheses and to guide future development:  

\begin{enumerate}
    \item How would you change the tool? Assume there’s no limit to the possible changes.
    \item What would you want us to keep the same about the tool?
    \item What aspects of the tool stood out and why?
\end{enumerate}

We ended with ten mixed demographic questions. We ended with the demographic questions to minimize the possibility of earlier results being impacted by stereotype threat. 

\begin{enumerate}
    \item ``What is your gender? (Short answer)''
    \item ``How old are you?'' With the options: 18-24, 25-35, 35-45, 45-55, and 55+
    \item ``Please pick the category of game design experience that best fits you:'' With the options: ``No experience (Never designed a game before)'', ``Limited game design experience (I've tried game development before)'', ``Regular game design experience (I've worked on multiple game projects before on my own or as a team)'', and ``I am a game design expert (Currently or in the past as part of daily life)''
    \item ``How would you describe your game design experience?'' (Short answer)
    \item  ``How often do you play games?'' With the options: Daily, A few times a week, Weekly, Monthly, an Less than monthly
    \item ``Pick the category of programming experience that best fits you:'' With the options: ``No experience (Never programmed)'', ``Limited programming experience (Have programmed before)'', ``Regular programming experience (Currently or in the past program weekly or monthly)'', and ``Programming expert (Currently or in the past program as part of daily life)''
    \item ``How would you describe your programming experience?'' (Short answer)
    \item ``Have you ever used a game design tool that did not require programming before? Ex) Scratch, Sony's Dreams, Project Spark, Game Builder Garage for Nintendo Switch, etc.'' With the options Yes and No.
    \item ``If yes to the above, what was the tool?'' (Short answer)
    \item ``Would you optionally want to provide an email address to be invited to a future study with an updated version of the tool?'' (Short answer)
\end{enumerate}

All but the last two questions were non-optional in order to ensure we collected the required results.

\subsection{Rule Example}
Figure \ref{fig:RuleBird} shows an example of a learned rule. Multiple of these rules create a learned mechanic. The top line shows the pre-effect of the bird's velocity moving down at a speed of -1, and then changing in the post-effect to a y velocity of +1. The first number in the values is the id of the object moving, in this case 0. All the lines below the pre and post effect are the conditions that must be true in order for the rule to fire. In this case the space bar has to be pressed, the longblock has to be moving left and the bird has to be falling.

\begin{figure}[tbh]
\centering
\begin{scriptsize}
\begin{verbatim}
RULE: 2 VelocityYFact: [0, -1.0]->VelocityYFact: [0, 1.0]
    VariableFact: ['space', True]
    VariableFact: ['up', False]
    VariableFact: ['down', False]
    VariableFact: ['left', False]
    VariableFact: ['right', False]
    VariableFact: ['upPrev', False]
    VariableFact: ['downPrev', False]
    VariableFact: ['leftPrev', False]
    VariableFact: ['rightPrev', False]
    VelocityYFact: [1, 0]
    VelocityXFact: [1, -1.0]
    AnimationFact: [1, 'longblock', 1.0, 4.0]
    VelocityXFact: [0, 0]
    VelocityYFact: [0, -1.0]
    AnimationFact: [0, 'bird', 1.0, 1.0]
\end{verbatim}
\caption{An example of a rule generated by the backend SMPS approach for the Flappy Bird game to make the bird jump.}
\label{fig:RuleBird}
\end{scriptsize}
\end{figure}

\end{document}